\documentclass[fleqn,usenatbib]{mnras}

\usepackage{newtxtext,newtxmath}

\usepackage[T1]{fontenc}

\let\VANthebibliography\thebibliography
\def\thebibliography{\DeclareRobustCommand{\VAN}[3]{##3}\VANthebibliography}

\usepackage{makecell}
\usepackage{amsmath}
\usepackage{booktabs}
\usepackage{lineno}
\usepackage{threeparttable}
\usepackage{makecell}
\usepackage{CJKutf8}
\usepackage[squaren]{SIunits}
\usepackage{comment}
\usepackage{xcolor}
\usepackage{multirow}
\usepackage{appendix}
\usepackage{longtable,booktabs}
\usepackage{float}
\usepackage{graphicx}	% Including figure files
\usepackage{amsmath}	% Advanced maths commands
\usepackage{amssymb}	% Extra maths symbols
\usepackage[squaren]{SIunits}
\usepackage[export]{adjustbox}

\title{The Self-organized Criticality Behaviors of Two New Parameters in SGR J1935+2154}
\author[S. Xiao et al.]{
Shuo Xiao,$^{1,2}$\thanks{E-mail: xiaoshuo@gznu.edu.cn}
Shuang-Nan Zhang,$^{3,4}$\thanks{E-mail: zhangsn@ihep.ac.cn}
Shao-Lin Xiong,$^{3}$\thanks{E-mail: xiongsl@ihep.ac.cn}
Ping Wang,$^{3}$
Xiu-Juan Li,$^{5}$
Ai-Jun Dong,$^{1,2}$
\newauthor
Qi-Jun Zhi,$^{1,2}$
Di Li$^{6,4}$\thanks{E-mail: dili@nao.cas.cn}
\\
$^{1}$School of Physics and Electronic Science, Guizhou Normal University, Guiyang 550001, People’s Republic of China\\
$^{2}$Guizhou Provincial Key Laboratory of Radio Astronomy and Data Processing, Guizhou Normal University, Guiyang 550001, People’s Republic of China\\
$^{3}$Key Laboratory of Particle Astrophysics, Institute of High Energy Physics, Chinese Academy of Sciences, Beijing 100049, China\\
$^{4}$University of Chinese Academy of Sciences, Chinese Academy of Sciences, Beijing 100049, China\\
$^{5}$School of Cyber Science and Engineering, Qufu Normal University, Qufu 273165, People’s Republic of China\\
$^{6}$CAS Key Laboratory of FAST, NAOC, Chinese Academy of Sciences, Beijing 100101, China\\
}

\date{Accepted XXX. Received YYY; in original form ZZZ}

\pubyear{2022}

\begin{document}
\begin{CJK}{UTF8}{gbsn}
\label{firstpage}
\pagerange{\pageref{firstpage}--\pageref{lastpage}}
\maketitle

\begin{abstract}
The minimum variation timescale (MVT) and spectral lag of hundreds of X-ray bursts (XRBs) from soft gamma-ray repeater (SGR) J1935+2154 were analyzed in detail for the first time in our recent work, which are important probes for studying the physical mechanism and radiation region. In this work, we investigate their differential and cumulative distributions carefully and find that they follow power-law models. Besides, the distributions of fluctuations in both parameters follow the Tsallis $q$-Gaussian distributions and the $q$ values are consistent for different scale intervals. Therefore, these results indicate that both parameters are scale-invariant, which provides new parameters for the study of self-organized criticality systems. Interestingly, we find that the $q$ values for MVT and spectral lag are similar with duration and fluence, respectively.

\end{abstract}


\section{Introduction}
The phenomenon of self-similarities widely exists in temporal and spatial scales in astronomical systems, including soft gamma repeaters (SGRs) \citep{cheng1996earthquake,gougucs1999statistical,chang2017scale,wei2021similar,sang2022statistical}, X-ray flares of afterglows of gamma-ray burst (GRB) \citep{wang2013self}, fast radio bursts (FRBs) \citep{wang2017sgr,wei2021similar,sang2022statistical}, solar flares \citep{goodman2020new}, pulsar glitches \citep{melatos2008avalanche}, etc (see \cite{2014ApJ...782...54A} and \cite{aschwanden201625} for a review). Interestingly, by discovering similar distributions of released energies for earthquakes and SGRs \citep{gougucs1999statistical}, it was suggested that SGRs originate from starquakes in magnetars \citep{duncan1992formation,thompson1995soft}. 

The Self-organized criticality (SOC) in slowly driven nonlinear dissipative systems was proposed by \cite{bak1987self}, \cite{aschwanden2012statistical} and \cite{aschwanden201625} to explain the phenomenon of self-similarities (i.e. power-law size distributions), the spatial dimension and the classical diffusion of the fractal-diffusive SOC system can be estimated through the power-law index for the parameters' length scales, durations, peak fluxes and fluences. 
On the other hand, the distribution for
the parameter fluctuations at different times in a SOC system was found to follow a Tsallis $q$-Gaussian
function, and the $q$ values are constant
for different scale intervals (i.e. scale invariance of the avalanche size differences) \citep{caruso2007analysis}. The power-law index of the parameter distribution and the $q$ value of the parameter difference distribution can be inter-convertible \citep{caruso2007analysis,celikoglu2010analysis}.

SGR J1935+2154 is one of the most active, and the only magnetar from which an X-ray burst (XRB) and a FRB (a bright ms-long radio burst) have been observed in association \citep{li2021hxmt,bochenek2020fast,mereghetti2020integral,younes2021broadband,ridnaia2021peculiar,tavani2021x}). Recently \cite{wei2021similar} and \cite{,sang2022statistical} reported that the XRBs originated from a SOC system through analyzing the parameters the duration, waiting time, fluence and flux, that is, these parameters are scale-invariant. Similarly, the scale invariance of these parameters is also found in other magnetars (e.g. SGR J1550–5418, \cite{wang2017sgr}; SGR 1806–20, \cite{cheng1996earthquake}; 1900+14, \cite{gougucs1999statistical}) and astronomical sources.

\begin{figure*}
\centering
\begin{minipage}[t]{0.48\textwidth}
\centering
\includegraphics[width=\columnwidth]{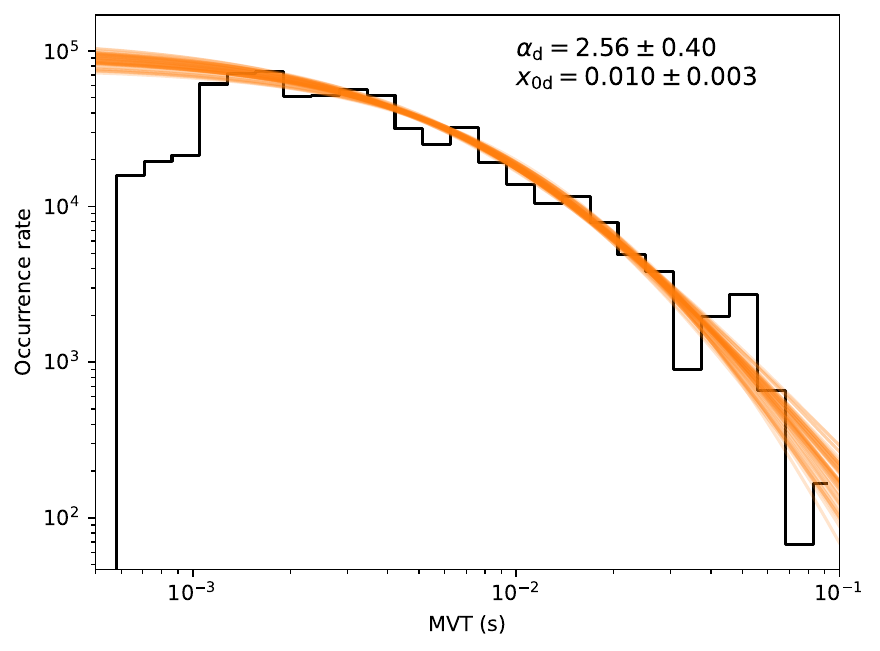}
\end{minipage}
\begin{minipage}[t]{0.48\textwidth}
\centering
\includegraphics[width=\columnwidth]{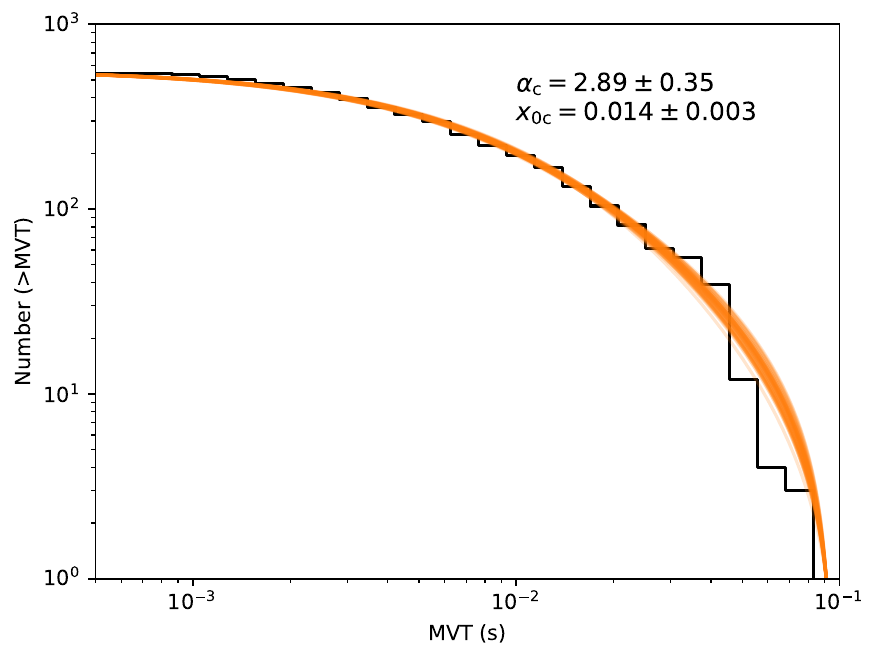}
\end{minipage}
\caption{The differential (left panel) and cumulative (right panel) distributions of MVTs observed by GECAM, HXMT and Fermi/GBM. The fitted yellow lines are obtained by MCMC with reduced-$\chi^2$ 1.9 and 0.6, respectively. }\label{mvt_dis}
\end{figure*}

\begin{figure*}
\centering
\begin{minipage}[t]{0.48\textwidth}
\centering
\includegraphics[width=\columnwidth]{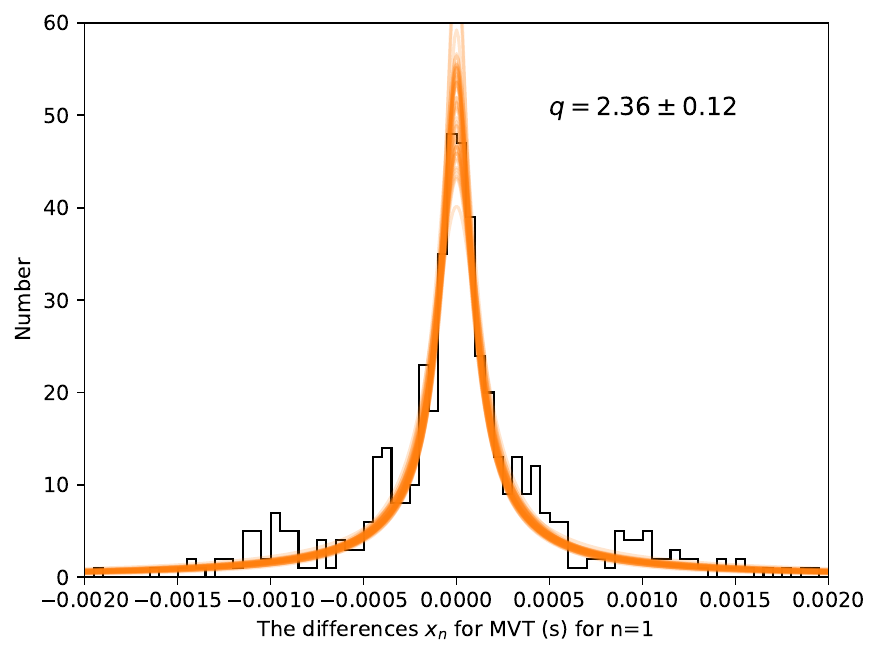}
\end{minipage}
\begin{minipage}[t]{0.48\textwidth}
\centering
\includegraphics[width=\columnwidth]{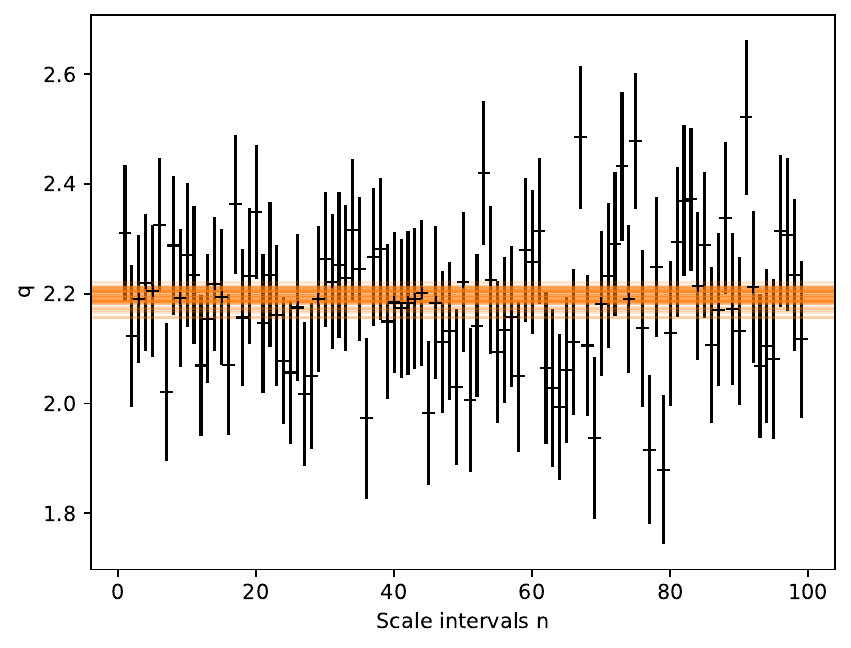}
\end{minipage}
\caption{Left panel: the distribution of fluctuations (MVT) for $n$=1, the yellow lines are fitted by a Tsallis $q$-Gaussian distribution ($q=2.36\pm0.12$), and reduced-$\chi^2$ is 0.8. Right panel: the fitted q values for different $n$, the yellow lines are fitted by a constant, which is $q=2.19\pm0.01$ with reduced-$\chi^2$ is 0.9.}\label{mvt_qs}
\end{figure*}

\begin{figure*}
\centering
\begin{minipage}[t]{0.48\textwidth}
\centering
\includegraphics[width=\columnwidth]{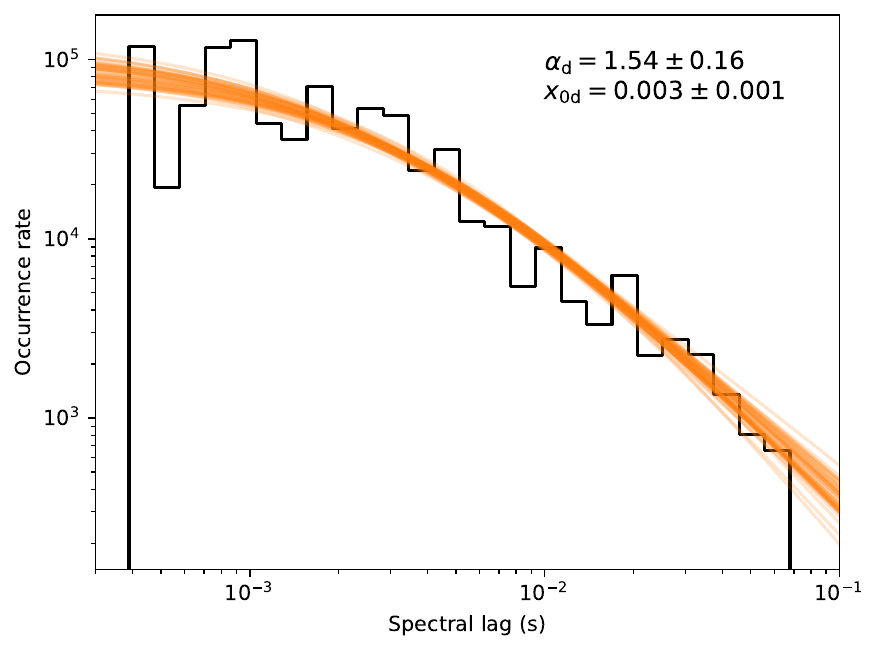}
\end{minipage}
\begin{minipage}[t]{0.48\textwidth}
\centering
\includegraphics[width=\columnwidth]{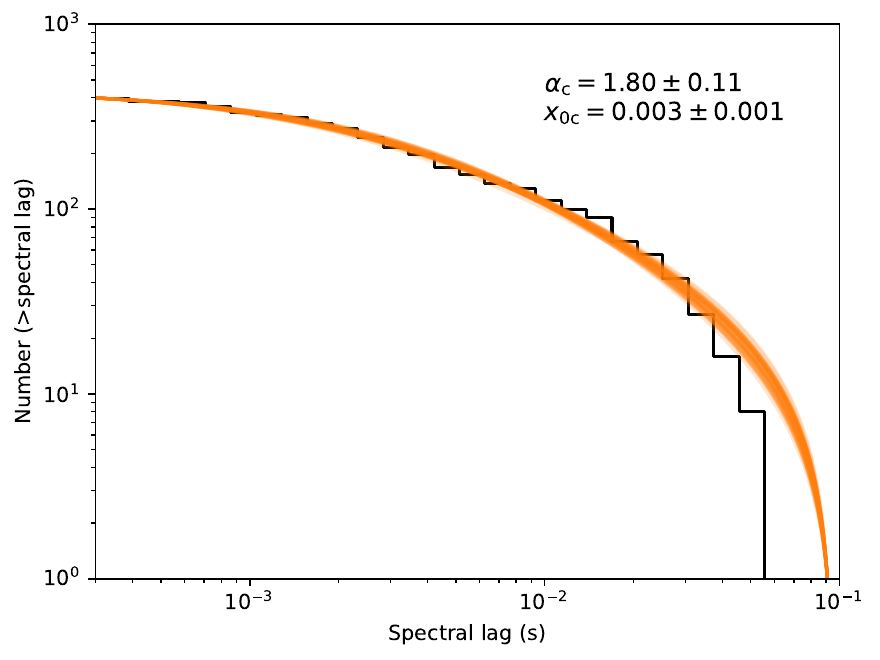}
\end{minipage}
\caption{The differential (left panel) and cumulative (right panel) distributions of spectral lags observed by GECAM, HXMT and Fermi/GBM. The fitted yellow lines are obtained by MCMC, and reduced-$\chi^2$ are 2.3 and 0.9, respectively.}\label{lag_dis}
\end{figure*}

\begin{figure*}
\centering
\begin{minipage}[t]{0.48\textwidth}
\centering
\includegraphics[width=\columnwidth]{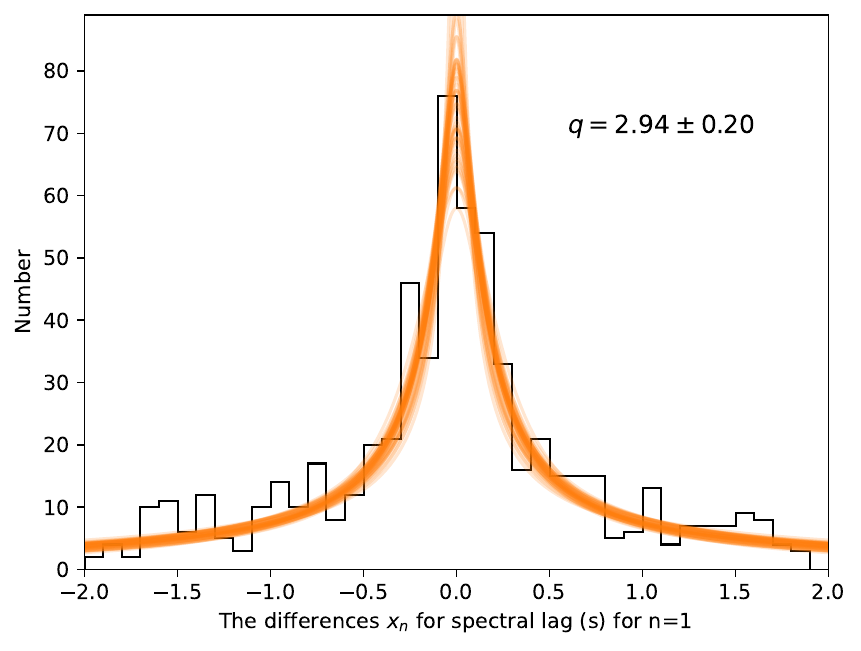}
\end{minipage}
\begin{minipage}[t]{0.48\textwidth}
\centering
\includegraphics[width=\columnwidth]{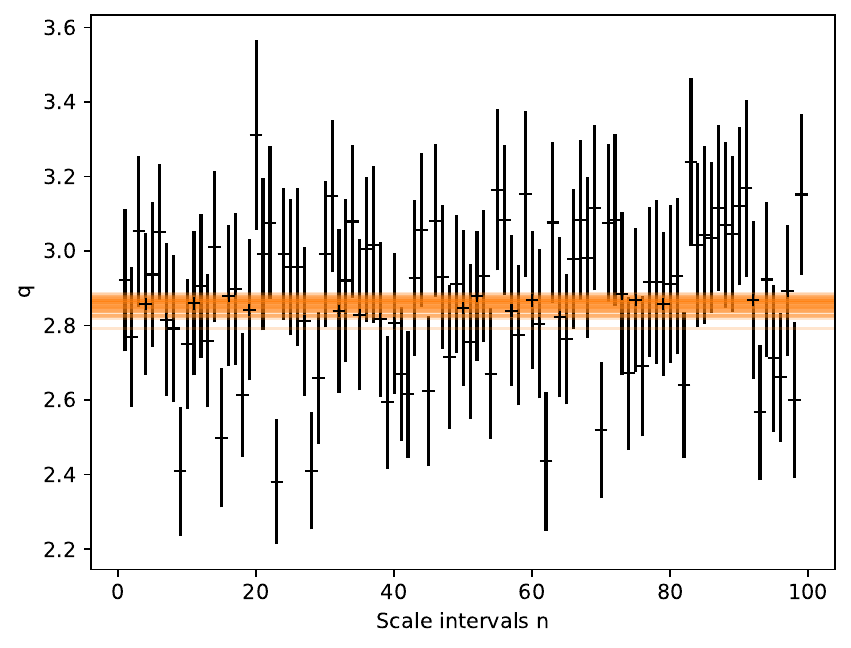}
\end{minipage}
\caption{Left panel: the distribution of fluctuations (spectral lag) for $n$=1, the yellow lines are fitted by a Tsallis $q$-Gaussian distribution ($q=2.94\pm0.20$), and reduced-$\chi^2$ is 1.5. Right panel: the fitted q values for different $n$, the yellow lines are fitted by a constant, which is $q=2.86\pm0.02$, and reduced-$\chi^2$ is 1.1.}\label{lag_qs}
\end{figure*}

In our recent work, we reported the small spectral lags (the distribution peaks $\sim$ 1.3 ms) \citep{xiao2023discovery} and the minimum variation timescale (MVT, the distribution peaks at $\sim$ 2 ms) \citep{xiao2023minimum} by analyzing hundreds of XRBs from SGR J1935+2154 observed by GECAM, HXMT and GBM. The two parameters are important probes to studying the physical mechanism and radiation region of XRBs \citep{1978Natur.271..525S, fenimore1993escape, 2007ApJ...660..556T,zhang2009discerning,huppenkothen2013quasi,ackermann2014fermi, golkhou2014uncovering,golkhou2015energy}. However, to our knowledge, whether these parameters are also scale-invariant has not been studied, which is the focus of this work.

In this work, we present our sample selection and scale invariance analysis in Section 2 and a discussion and summary are given in Section 3.

\section{Scale invariance analysis}

\subsection{Sample selection and method}
The bursts from SGR J1935+2154 and the values of MVT and spectral lag observed by GECAM, HXMT and GBM are collected from our previous work \citep{xiao2023minimum,xiao2023discovery}. Since all three satellites have large fields of view and high sensitivity \citep{xiong2020gecam,zhang2020overview,2009ApJ...702..791M}, we have a relatively comprehensive sample. For bursts observed jointly by multiple satellites, we adopt the GBM measurements by default. In total, 669 bursts are selected from July 2014 to January 2022 after removing those with “timing glitches” (a known GBM hardware anomaly with dips and peaks in a light curve \citep{briggs2013terrestrial}.) or saturation.

To investigate whether the parameters are scale-invariant or not, we adopt two approaches like Refs. \citep{chang2017scale,wei2021similar,sang2022statistical,li2023evidence}. The first is to investigate whether the differential and cumulative distributions of the parameters follow power-law models \citep{aschwanden2015thresholded} (see Equations \ref{pl0} and \ref{pl1}) and the power-law index are consistent within the uncertainties, 
\begin{equation} \label{pl0}
\begin{split}
n_0=N(1-\alpha_{\rm d})[(x_2+x_0)^{(1-\alpha_{\rm d})}-(x_1+x_0)^{(1-\alpha_{\rm d})}]^{-1} , \\
N_{\rm diff}=n_0(x_{\rm 0d}+x)^{-\alpha_{\rm d}},
\end{split}
\end{equation}
\begin{equation} \label{pl1}
N_{\rm cum}(>x)=1+(N-1) \times \frac{(x_2+x_{\rm 0c})^{1-\alpha_{\rm c}}-(x+x_{\rm 0c})^{1-\alpha_c}}{(x_2+x_{\rm 0c})^{1-\alpha_c }-(x_1+x_{\rm 0c})^{1-\alpha_{\rm c}}},
\end{equation}
where $n_0$, $\alpha_{d}$ and $\alpha_{c}$ are normalization constant, the power-law index of differential and cumulative distributions, respectively. $N$ is the number of bursts, $x_{0d}$ and $x_{0c}$ are constants due to the threshold effects.

The second is to test whether the distributions of the differences of the parameters $X_n$ follow the Tsallis $q$-Gaussian distribution \citep{1988JSP....52..479T,1998PhyA..261..534T} (Equation \ref{eq:qGauss}) and the $q$ values for different time scales $n$ are consistent for different scale intervals,
\begin{equation}
X_n=S_{i+n}-S_{i},
\label{Xn}
\end{equation}
where $S_i$ is the value of the parameter (i.e. MVT and spectral lag) of the $i$th burst ordered according to time and $n$ is the temporal interval scale. We obtain $x_n$ by dividing the difference $X_n$ by a scale factor ($\sigma_{X_n}$), which is the standard deviation of $X_n$.
\begin{equation}
f(x_{n})=\alpha\left[1-\left(1-q\right)x_{n}^{2}/\beta\right]^{\frac{1}{1-q}},
\label{eq:qGauss}
\end{equation}
where $\alpha$ is the normalization factor, the $\beta$ and $q$ (the parameter we are interested in) affect the width and sharpness of the peak, respectively. We adopted the Markov Chain Monte Carlo (MCMC) method to fit the data, and tests the goodness-of-fit by reduced-$\chi^2$ (i.e. close to 1 implies a good fit). 

\subsection{Minimum variation timescales}
The MVT in 8-100 keV of each burst is calculated in \cite{xiao2023minimum}. The left panel of Figure~\ref{mvt_dis} shows the differential distribution, which can be fitted fairly well  (reduced-$\chi^2$=$\chi^2/dof$=37.4/19=1.9) by the power-law model (Equation \ref{pl0}) with $\alpha_{\rm d}=2.56\pm0.40$ and $x_{\rm 0d}=0.010\pm0.003$. Note that less than $\sim$0.001 ms does not fit well due to threshold effects \citep{aschwanden2015thresholded}. The cumulative distribution of MVT is shown in the right panel of Figure~\ref{mvt_dis}, which also can be fitted well (reduced-$\chi^2$=$\chi^2/dof$=12.6/23=0.6) by the power-law model (Equation \ref{pl1}) with $\alpha_{\rm c}=2.89\pm0.35$ and $x_{\rm 0c}=0.014\pm0.003$. Note that $\alpha_{\rm d}$ and $\alpha_{\rm c}$ are consistent within the uncertainties, which indicates that the MVT parameter of XRBs from SGR J1935+2154 exhibits SOC behavior.

The distributions of fluctuations for different time scales $n$ are investigated, for example, the left panel of Figure~\ref{mvt_qs} shows the distribution for $n$=1, and it can be fitted well (reduced-$\chi^2$=$\chi^2/dof$=58.1/76=0.8) by the Tsallis $q$-Gaussian distribution (Equation \ref{eq:qGauss}) with $q=2.36\pm0.12$. The q-values obtained by fitting the distributions for different scales $n$ are shown in the right panel of Figure~\ref{mvt_qs}, which are steady and can be fitted well with a constant ($q=2.19\pm0.01$, with reduced-$\chi^2$=$\chi^2/dof$=83.3/98=0.9), which also verifies that the MVT parameter of XRB from SGR J1935+2154 exhibits SOC behavior.

\subsection{Spectral lags}
The spectral lag between light curves in the 10-20 and 60-100 keV energy bands is calculated in \cite{xiao2023discovery}, and a positive lag is defined as that low-energy photons follow high-energy photons. The differential and cumulative distributions are shown in Figure~\ref{lag_dis}, both follow power-law models well. The result of the fit to the difference distribution is $\alpha_{\rm d}=1.54\pm0.16$ and $x_{\rm 0d}=0.003\pm0.001$ with reduced-$\chi^2$=$\chi^2/dof$=43.9/19=2.3. For the cumulative distribution, the result of the fit is $\alpha_{\rm c}=1.80\pm0.11$ and $x_{\rm 0c}=0.003\pm0.001$, with reduced-$\chi^2$=$\chi^2/dof$=11.5/13=0.9. Note that the $\alpha_{\rm d}$ and $\alpha_{\rm c}$ are also consistent within the uncertainties, which indicates that the spectral lag parameter of XRB from SGR J1935+2154 exhibits SOC behavior.

The left panel of Figure~\ref{lag_qs} shows the distribution of fluctuations for time scale $n=1$, which can be fitted well (reduced-$\chi^2$=$\chi^2/dof$=52.3/36=1.5) by the Tsallis $q$-Gaussian distribution (Equation \ref{eq:qGauss}) with $q=2.94\pm0.20$. The q-values obtained by fitting the distributions for different scales $n$ are also consistent within the uncertainties and fitted with a constant ($q=2.86\pm0.02$, reduced-$\chi^2$=$\chi^2/dof$=103.7/98=1.1), which also verifies that the spectral lag parameter of XRB from SGR J1935+2154 exhibit SOC behavior.

\section{Discussion and conclusion}
In this work, we perform a detailed study of the statistical properties of the two parameters, i.e., MVT and spectral lag, of hundreds of XRBs from SGR J1935+2154 observed by GECAM, HXMT and GBM. With two methods, we report that both parameters exhibit SOC behaviors, that is, the differential and cumulative distributions of the parameters well follow power-law models and the power-law index ($\alpha_{\rm d}$ and $\alpha_{\rm c}$) are consistent within the uncertainties, as well as the distributions of fluctuations follow the Tsallis $q$-Gaussian distribution and the $q$ values for different time scales are consistent for different scale intervals.

Interestingly, the $q$ values (i.e. $q=2.19\pm0.01$) for MVT are similar to the result (i.e. $2.28\pm 0.15$) for the duration reported by \cite{wei2021similar}. The duration of a burst only describes the total emission properties, but MVT captures the information concerning individual pulses, that is, the MVT is approximately equal to the rise time of the shortest pulse in a burst. In a previous work, we did not find a significant correlation between duration and MVT for XRBs from SGR J1935+2154 \citep{xiao2023minimum}. The MVT of XRBs can be used to estimate the emission region in pulsar-like models, as well as the radius and Lorentz factor of the relativistic jet in GRB-like models (see \citealp{2020Natur.587...45Z,2022arXiv221203972Z} for review). Therefore, the SOC behavior of MVT implies that the radius of the radiation region may have a similar origin in pulsar-like models.

On the other hand, although the spectral lag is a time-domain parameter, the $q$ values (i.e. $q=2.86\pm0.02$) for spectral lag are similar to the result (i.e. $2.78\pm 0.12$) for fluence reported by \cite{wei2021similar}. The relationship between fluence and spectral lag has been found in some transient sources such as GRBs \citep{ukwatta2012lag}, but there is a lack of studies for XRBs. Further work is required to investigate this.
 
According to the study of \cite{caruso2007analysis} and \cite{celikoglu2010analysis}, there is an exact relation (i.e. $q=(\alpha+2)/2$) between the power-law index ($\alpha_{\rm d}$ or $\alpha_{\rm c}$) and the $q$ values based on the assumption that the sizes of two events have no correlation. We find that this relationship is also satisfied for MVT and spectral lag within the uncertainties. However, MVT and spectral lag are different from other parameters, such as duration, peak flux, and fluence or energy, which have detailed theoretical explanations, e.g.  the spatial dimension can be estimated based on the Fractal-diffusive Avalanche Model \citep{aschwanden2012statistical}. Therefore, we believe that the SOC behaviours of MVT and spectral lag will provide new insights in future SOC studies of these phenomen.

\section*{Acknowledgments}
We acknowledge the public data from {\it Fermi}/GBM. This work is supported by the National Natural Science Foundation of China (NOs. 12303043, 12273042 and 12273008) and the National Key R\&D Program of China 2022YFF0711404.
The authors also thank supports from 
the Strategic Priority Research Program on Space Science, the Chinese Academy of Sciences (Grant No.
XDA15360300, %% GECAM science appliacation
XDA15052700), %% Xiong shaolin, GECAM data analysis
the National SKA Program of China (Nos. 2022SKA0130100, 2022SKA0130104), the Natural Science and Technology Foundation of Guizhou Province (No. [2023]024) and the Foundation of Guizhou Provincial Education Department (No. KY (2020) 003).

% \newpage
\section*{DATA AVAILABILITY}
The data underlying this article will be shared on reasonable request
to the corresponding author.
% \newpage
\bibliographystyle{mnras}
\bibliography{ref0} % if your bibtex file is called example.bib
% \newpage

% Don't change these lines
\bsp	% typesetting comment
% \label{lastpage}
\end{CJK}
\end{document}